\documentclass[aps,preprint,showpacs,preprintnumbers,eqsecnum,amsmath,amssymb]{revtex4}
\usepackage{graphics,epsfig}
\usepackage{graphicx}
\usepackage{dcolumn}
\usepackage{bm}

\begin{document}

\title{Dirac Quasinormal frequencies of the Kerr-Newman black hole}
\author{Jiliang Jing} \email{jljing@hunnu.edu.cn}
\author{Qiyuan Pan}
\affiliation{ Institute of Physics and  Department of Physics, \\
Hunan Normal University,\\ Changsha, Hunan 410081, P. R. China }

\baselineskip=0.64 cm

\vspace*{0.2cm}
\begin{abstract}
\vspace*{0.2cm}

\baselineskip=0.64 cm

The Dirac quasinormal modes (QNMs) of the Kerr-Newman black hole
are investigated using continued fraction approach. It is shown
that the quasinormal frequencies in the complex $\omega$ plane
move counterclockwise as the charge or angular momentum per unit
mass of the black hole increases. They get a spiral-like shape,
moving out of their Schwarzschild or Reissner-Nordstr\"om values
and ``looping in" towards some limiting frequencies as the charge
and angular momentum per unit mass  tend to their extremal values.
The number of the spirals increases as the overtone number
increases but decreases as the angular quantum number increases.
It is also found that both the real and imaginary parts are
oscillatory functions of the angular momentum per unit mass, and
the oscillation becomes faster as the overtone number increases
but slower as the angular quantum number increases.

\end{abstract}

 \vspace*{1.5cm}


 \maketitle

\section{introduction}

It is well known that QNMs possess discrete spectra of complex
characteristic frequencies which are entirely fixed by the
structure of the background spacetime and irrelevant of the
initial perturbations\cite{Chand75} \cite{Frolov98}. Thus, it is
believed that one can directly identify the existence of a black
hole by comparing QNMs with the gravitational waves observed in
the universe, as well as test the stability of the event horizon
against small perturbations. Meanwhile, it is generally believed
that the study of QNMs may lead to a deeper understanding of the
thermodynamic properties of black holes in loop quantum gravity
\cite{Hod} \cite{Dreyer} since the real part of quasinormal
frequencies with a large imaginary part for the scalar field in
the Schwarzschild black hole is equal to the Barbero-Immirzi
parameter \cite{Hod, Dreyer, Baez, Kunstatter}, a factor
introduced by hand in order that the loop quantum gravity
reproduces correct entropy of the black hole. It was also argued
that the QNMs of Anti-de Sitter (AdS) black holes have a direct
interpretation in terms of the dual conformal field theory (CFT)
\cite{Maldacena, Witten, Kalyana}. According to the AdS/CFT
correspondence, a large static black hole in asymptotically AdS
spacetime corresponds to a (approximately) thermal state in CFT,
and the decay of the test field in the black-hole spacetime
corresponds to the decay of the perturbed state in CFT. The
dynamical timescale for the return to the thermal equilibrium can
be done in the AdS spacetime, and then translated into the CFT
using the AdS/CFT correspondence. Thus, QNMs for various  black
holes have been studied extensively \cite{Chan}-\cite{Kurita}.

Although many authors studied the QNMs of scalar, electromagnetic
and gravitational perturbations, the investigation of the Dirac
QNMs is very limited \cite{Cho, Shen, Hu, Zhidenko, Jing1}.
Furthermore, in these papers the study of the Dirac QNMs was
limited to use P\"oshl-Teller potential or WKB approximative
methods. The P\"oshl-Teller potential approximative may be a
valuable tool in certain situations, but in general results
obtained from the approximating potentials should not be expected
to be accurate. The WKB method can be generalized to deal with the
Reissner-Nordstr\"om and Kerr black holes. In the latter case,
unfortunately, the method is not very simple to apply and does not
yield very accurate results. It is well known that the Leaver's
continued fraction technique is the best workhorse method to
compute highly damped modes and is very reliable. However, the
method can not be directly used to study the QNMs of the Dirac
fields if we take the standard wave equations. The reason to
obstruct the study of the Dirac QNMs using continued fraction
technique is that for static black holes the standard wave
equation
\begin{eqnarray}\label{sta}
\left(\frac{d^2}{d r_*^2}+\omega^2\right)Z_{\pm}=V_{\pm}Z_{\pm}
\end{eqnarray}
possesses a special potential
 \begin{eqnarray}
V_{\pm}=\lambda^2\frac{ \Delta }{r^4}\pm \lambda \frac{d}{d
r_*}\frac{\sqrt{\Delta}}{r^2} \label{V00}
 \end{eqnarray}
which is the function of $\sqrt{\Delta}$
\cite{Cho}\cite{Jing1}\cite{Jing2}, where $\Delta$ is a function
related to the metric of the spacetimes, say $\Delta=r^2-2Mr$ for
the Schwarzschild black hole. We can not investigate the Dirac
QNMs using the continued fraction approaches because we have to
expand the potential as a series at the event horizon but the
factor $\sqrt{\Delta}$ does not permit us to do that.

Recently, we \cite{Jing3, Jing4} found that the wave function and
the potential of the Dirac fields can be expressed as new forms.
Starting from the new wave function and potential, we \cite{Jing3,
Jing4} studied the Dirac QNMs of the Schwarzschild black hole
using the continued fraction \cite{Leaver} and Hill-determinant
approaches \cite{Majumdar}\cite{Leaver2} and the Dirac QNMs of the
Schwarzschild-anti-de Sitter and Reissner-Nordstr\"om-anti-de
Sitter black holes using Horowitz-Hubeny approach \cite{Horowitz}.
The main purpose of this paper is to extend the investigation to
the Kerr-Newman black hole because this black hole is the only
asymptotically flat electro-vacuum solution of the
Einstein-Maxwell system \cite{Kerr} and it is the most general
black hole among classical black holes.

The organization of this paper is as follows. In Sec. 2 the
decoupled Dirac equations and corresponding wave equations in the
Kerr-Newman spacetime are obtained using Newman-Penrose formalism.
In Sec. 3 the angular and radial continued fraction equations are
introduced. In Sec. 4 the numerical results for the Dirac QNMs in
the Reissner-Nordstr\"om, Kerr and Kerr-Newman black holes are
presented. The last section is devoted to a summary.

\section{Dirac equations in the Kerr-Newman spacetime}

In a curve spacetime the Dirac equations \cite{Page} can be
expressed as
\begin{eqnarray}
   &&\sqrt{2}\nabla_{BB'}P^B+i\mu \bar{Q}_{B'}=0, \nonumber \\
   &&\sqrt{2}\nabla_{BB'}Q^B+i\mu \bar{P}_{B'}=0,
\end{eqnarray}
where $\nabla_{BB'}$ is covariant differentiation, $P^B$ and $Q^B$
are the two-component spinors representing the wave functions, $
\bar{P}_{B'}$ is the complex conjugate of $P_{B}$, and $\mu $ is
the  mass of the Dirac particle. In the Newman-Penrose formalism
\cite{Newman} the equations become
\begin{eqnarray}\label{np}
   &&(D+\epsilon-\rho )P^0+
   (\bar{\delta}+\pi-\alpha )P^1=2^{-1/2}i\mu  \bar{Q}^{1'},\nonumber \\
   &&(\triangle+\mu -\gamma )P^1+
   (\delta+\beta -\tau )P^0=-2^{-1/2}i\mu
    \bar{Q}^{0'}, \nonumber\\
   &&(D+\bar{\epsilon}-\bar{\rho} )\bar{Q}^{0'}+
   (\delta+\bar{\pi}-\bar{\alpha} )\bar{Q}^{1'}=-2^{-1/2}i\mu  P^{1},\nonumber \\
  &&(\triangle+\bar{\mu} -\bar{\gamma} )\bar{Q}^{1'}+
   (\bar{\delta}+\bar{\beta} -\bar{\tau} )\bar{Q}^{0'}=2^{-1/2}i\mu
   P^{0}.
\end{eqnarray}
The null tetrad for the Kerr-Newman black hole can be taken as
\begin{eqnarray}
  &&l^\mu=\left(\frac{r^2+a^2}{\Delta}, ~1, ~0, ~\frac{a}{\Delta} \right), \nonumber \\
  &&n^\mu=\frac{1}{2\rho\bar{\rho}}((r^2+a^2), ~-\Delta, ~0, ~a),\nonumber \\
  &&m^\mu=\frac{1}{\sqrt{2} \bar{\rho}}\left(i a sin\theta, ~0, ~1, \frac{i}{sin\theta}\right),
\end{eqnarray}
with
\begin{eqnarray}
&&\rho=r-i a cos\theta, \nonumber \\
&&\Delta=r^2-2M r+a^2+Q^2=(r-r_+)(r-r_-),
\end{eqnarray}
where $M$, $Q$ and $a$ represent the mass, charge and angular
momentum per unit mass  of the Kerr-Newman black hole
respectively, and $r_{\pm}=M\pm \sqrt{M^2-a^2-Q^2}$. As
$Q\rightarrow 0$ the Kerr-Newman metric reduces to the rotating
Kerr metric, and as $a\rightarrow 0$ it reduces to the charged
Reissner-Nordstr\"om metric.  Then the wave functions can be taken
as
\begin{eqnarray}
&&P^0=\frac{1}{\rho}{\mathbb{R}}_{-1/2}(r)S_{-1/2}(\theta)
e^{-i(\omega t-m\varphi)}, \nonumber \\
&&P^1={\mathbb{R}}_{+1/2}(r)S_{+1/2}(\theta)
e^{-i(\omega t-m\varphi)}, \nonumber \\
&&\bar{Q}^{1'}={\mathbb{R}}_{+1/2}(r)S_{-1/2}(\theta)
e^{-i(\omega t-m\varphi)}, \nonumber \\
&&\bar{Q}^{0'}=-\frac{1}{\bar{\rho}}{\mathbb{R}}_{-1/2}(r)S_{+1/2}(\theta)e^{-i(\omega
t-m\varphi)},
\end{eqnarray}
where $\omega$ and $m$ are the energy and angular momentum of the
Dirac particle. After the tedious calculation Eq. (\ref{np}) can
be simplified as
\begin{eqnarray}\label{dd2}
&&\sqrt{\Delta}{\mathcal{D}}_0 {\mathbb{R}}_{-1/2}=(\lambda+i\mu
r)\sqrt{\Delta} {\mathbb{R}}_{+1/2}, \\
\label{dd3}&&\sqrt{\Delta}{\mathcal{D}}_0^{\dag}
(\sqrt{\Delta}{\mathbb{R}}_{+1/2})=(\lambda-i\mu r) {\mathbb{R}}_{-1/2},\\
&&{\mathcal{L}}_{1/2} S_{+1/2}=-(\lambda-a \mu cos\theta )S_{-1/2}, \label{aa1}\\
&&{\mathcal{L}}_{1/2}^{\dag} S_{-1/2}=(\lambda+a \mu cos\theta)
S_{+1/2}.\label{aa2}
\end{eqnarray}
with
 \begin{eqnarray}
 &&{\mathcal{D}}_n=\frac{\partial}{\partial r}-\frac{i K}
 {\Delta}+\frac{n}{\Delta}\frac{d \Delta}{d r},\nonumber \\
 &&{\mathcal{D}}^{\dag}_n=\frac{\partial}{\partial r}+\frac{i K}
 {\Delta}+\frac{n}{\Delta}\frac{d \Delta}{d r},\nonumber \\
 &&{\mathcal{L}}_n=\frac{\partial}{\partial \theta}-a \omega sin\theta
 +\frac{m}{\sin \theta }
 +n\cot \theta,\nonumber \\
 &&{\mathcal{L}}^{\dag}_n=\frac{\partial}{\partial \theta}
 +a \omega sin\theta-\frac{m}{\sin \theta }
 +n\cot \theta, \nonumber \\
 &&K=(r^2+a^2)\omega-m a.\label{ld}
 \end{eqnarray}

We can eliminate $S_{+1/2}$ (or $S_{-1/2}$) from Eqs. (\ref{aa1})
and (\ref{aa2}). Defining $u=cos \theta$, we find that the angular
equation can be expressed as
\begin{eqnarray}\label{ang}
\frac{d}{du}\left[(1-u^2)\frac{d S_{s}}{d u}\right]+\left[
(a\omega u)^2-2 a \omega s u +s
+A_{lm}-\frac{(m+su)^2}{1-u^2}\right]S_s=0,
\end{eqnarray}
where $A_{lm}$ is the angular separation constant. It is
interesting to note that the angular equation (\ref{ang}) is the
same as in the Kerr case.

We will focus our attention on the massless Dirac field in this
paper. Therefore, we can eliminate ${\mathbb{R}}_{-1/2}$ (or
$\sqrt{\Delta}{\mathbb{R}}_{+1/2}$) from Eqs. (\ref{dd2}) and
(\ref{dd3}) to obtain a radial decoupled Dirac equation for
$\sqrt{\Delta} {\mathbb{R}}_{+1/2}$ (or ${\mathbb{R}}_{-1/2}$).
Then, introducing an usual tortoise coordinate
 \begin{eqnarray}
dr_*=\frac{r^2+a^2}{\Delta} dr,
  \end{eqnarray}
 and resolving the equation in the form
 \begin{eqnarray} \label{Rphi}
&&{\mathbb{R}}_{s}=\frac{\Delta^{-s/2}}{\sqrt{r^2+a^2}} \Psi_s,
 \end{eqnarray}
we obtain the wave equation
\begin{eqnarray}\label{wave}
\frac{d^2 \Psi_s }{d r_*^2}+(\omega ^2-V_s )\Psi_s =0,
\end{eqnarray}
with
\begin{eqnarray}\label{Poten}
V_s=-\frac{d Z}{d r_*}+Z^2+\frac{\Delta}{(r^2+a^2)^2}\left[\frac{2
m a \omega(r^2+a^2)-m^2a^2}{\Delta}+\frac{i s
 K}{\Delta}\frac{d \Delta}{d r}-4 i s \omega r +\lambda^2\right],
\end{eqnarray}
where $Z=-\frac{s}{2(r^2+a^2)}\frac{d \Delta}{dr}-\frac{r
\Delta}{(r^2+a^2)^2}$ and $\lambda^2=A_{lm}-2m a \omega +(a
\omega)^2$. We will see that we can easily work out the Dirac
quasinormal frequencies of the Kerr-Newman black hole from Eqs.
(\ref{ang}) and (\ref{wave}) using the continued fraction
approach.

\section{Angular and radial continued fraction equations}

For the Kerr-Newman black hole, we have to solve Eq. (\ref{ang})
numerically following Leaver \cite{Leaver}.  Boundary conditions
for Eq. (\ref{ang}) are that $S_s$ is regular at the regular
singular points $u=\pm 1$ where the indices are given by $\pm
(m+s)/2$ at $u=1$ and $\pm (m-s)/2$ at $u=-1$. So a solution to
Eq. (\ref{ang}) may be expressed as
\begin{equation}
\label{6} S_s(u)=e^{a\omega u}(1+u)^{k_1}(1-u)^{k_2}
\sum^{\infty}_{n=0}a^{\theta}_{n}(1+u)^n,
\end{equation}
where $k_1=|m-s|/2$, $k_2=|m+s|/2$, and the superscript $\theta$
denotes the association with the angular equation. The series
coefficients are related by a three term recurrence relation and
the boundary condition at $u=+1$ is satisfied only by its minimal
solution sequence. The three-term recurrence relation can be
written as
\begin{eqnarray}
\label{7} &&
\alpha^{\theta}_{0}a^{\theta}_{1}+\beta^{\theta}_{0}a^{\theta}_{0}=0,\\
\label{8} &&
\alpha^{\theta}_{n}a^{\theta}_{n+1}+\beta^{\theta}_{n}a^{\theta}_{n}
                     +\gamma^{\theta}_{n}a^{\theta}_{n-1}=0,~~~~~~~~ (n\geq 1),
\end{eqnarray}
where
\begin{eqnarray}
\alpha^{\theta}_{n}&=&-2(n+1)(n+2k_1+1),\nonumber \\
\beta^{\theta}_{n}&=&n(n-1)+2n(k_1+k_2+1-2a\omega)-2a\omega(2k_1+s+1)
              \nonumber \\ &&+(k_1+k_2)(k_1+k_2+1)-[a^2\omega^2+s(s+1)+A_{lm}],\nonumber \\
\gamma^{\theta}_{n}&=&2a\omega(n+k_1+k_2+s).
\end{eqnarray}
We will obtain the minimal solution if the angular separation
constant $A_{lm}$ is a root of the continued fraction equation
\begin{equation}\label{12}
0=\beta^{\theta}_{0}-\frac{\alpha^{\theta}_{0}\gamma^{\theta}_{1}}
{\beta^{\theta}_{1}-}\frac{\alpha^{\theta}_{1}\gamma^{\theta}_{2}}
{\beta^{\theta}_{2}-}\frac{\alpha^{\theta}_2\gamma^{\theta}_3}
{\beta^{\theta}_{3}-}\frac{\alpha^{\theta}_3\gamma^{\theta}_4}{\beta^{\theta}_{4}-}\cdots.
\end{equation}

The QNMs of the Kerr-Newman black hole are defined to be the modes
with purely ingoing waves at the event horizon and purely outgoing
waves at infinity \cite{Chand75}. Then, the boundary conditions on
wave function $\Psi_s$ at the horizon $(r=r_+)$ and infinity
$(r\rightarrow +\infty)$ can be expressed as
 \begin{eqnarray}
 \label{Bon}
\Psi_s  \sim \left\{
\begin{array}{ll} (r-r_+)^{-\frac{s}{2}-i \sigma_+} &
~~~~r\rightarrow r_+, \\
     r^{-s+i\omega}e^{i\omega r} & ~~~~     r\rightarrow +\infty,
\end{array} \right.
 \end{eqnarray}
where $\sigma_\pm=\frac{1}{r_+-r_-}\left[(r^2_\pm +a^2)\omega -m
a\right]$.

A solution to Eq. (\ref{wave}) which satisfies the desired
behavior at the boundary can be written in the form
 \begin{eqnarray}\label{expand}
 \Psi_s=\sqrt{r^2+a^2} (r-r_+)^{-\frac{s}{2}
 -i \sigma_+}(r-r_-)^{-1-\frac{s}{2}+2i\omega
 +i\sigma _-}e^{i\omega (r-r_-)}\sum_{n=0}^{\infty}
 a_n\left(\frac{r-r_+}{r-r_-}\right)^n.
 \end{eqnarray}
If we take $r_++r_-=1$ and $b=r_+-r_-$, the sequence of the
expansion coefficients $\{a_n: n=1,2,....\}$ is determined by a
three-term recurrence relation staring with $a_0=1$:
 \begin{eqnarray} \label{rec}
 &&\alpha_0 a_1+\beta_0 a_0=0, \nonumber \\
 &&\alpha_n a_{n+1}+\beta_n a_n+\gamma_n
 a_{n-1}=0,~~~(n=1,2,...).
 \end{eqnarray}
The recurrence coefficient $\alpha_n$, $\beta_n$ and $\gamma_n$
are given in terms of $n$ and the black hole parameters by
\begin{eqnarray}
 &&\alpha_n=n^2+(C_0+1)n+C_0, \nonumber \\
 &&\beta_n=-2n^2+(C_1+2)n+C_3, \nonumber  \\
 &&\gamma_n=n^2+(C_2-3)n+C_4-C_2+2,
 \end{eqnarray}
and the intermediate constants $C_n$ are defined by
\begin{eqnarray}
 C_0&=&1-s-i\omega-\frac{2 i}{b}\left[\left(\frac{r_+^2+r_-^2}{2}
 +a^2\right)\omega -ma\right], \nonumber \\
 C_1&=&-4+2i\omega(2+b)+\frac{4 i}{b}\left[\left(\frac{r_+^2+r_-^2}{2}
 +a^2\right)\omega -ma\right],  \nonumber  \\
 C_2&=&s+3-3i\omega-\frac{2 i}{b}\left[\left(\frac{r_+^2+r_-^2}{2}
 +a^2\right)\omega -ma\right], \nonumber \\
 C_3&=&\omega^2(4+2b-4r_+r_-+3 a^2)-2m a \omega
 -s-1+(2+b)i\omega-A_{lm}\nonumber \\
 &&+\frac{2(2\omega+ i)}{b}\left[\left(\frac{r_+^2+r_-^2}{2}
 +a^2\right)\omega -ma\right], \nonumber \\
 C_4&=&s+1-2\omega^2-(2s+3)i\omega-\frac{2(2\omega+ i)}{b}
 \left[\left(\frac{r_+^2+r_-^2}{2}
 +a^2\right)\omega -ma\right].
 \end{eqnarray}

It is interesting to note that the three-term recursion relation
is  obtained form the wave equation (\ref{wave}) directly. The
calculation is simpler than other cases. For example, the
coefficients of the expansion for the electromagnetic and
gravitational fields in the Reissner-Nordstr\"om black hole are
determined by a four-term recursion relation which should be
reduced to a three-term relation using a Gaussian eliminated step
(see \cite{Leaver2} for details). Our results are also more
concise than that for the electromagnetic and gravitational
perturbations in the Kerr-Newman black-hole spacetime obtained by
Berti and Kokkotas \cite{Berti}.

The series (\ref{expand}) converges and the $r=+\infty$ boundary
condition (\ref{Bon}) is satisfied if, for a given $s$, $a$, $m$
and $A_{lm}$, the frequency $\omega$ is a root of the continued
fraction equation
\begin{eqnarray}\label{ann}
 \left[\beta_n-\frac{\alpha_{n-1}\gamma_n}{\beta_{n-1}-}
 \frac{\alpha_{n-2}\gamma_{n-1}}{\beta_{n-2}-}...
 \frac{\alpha_0\gamma_1}{\beta_0}\right]=
 \left[\frac{\alpha_n\gamma_{n+1}}{\beta_{n+1}-}
 \frac{\alpha_{n+1}\gamma_{n+2}}{\beta_{n+2}-}
 \frac{\alpha_{n+2}\gamma_{n+3}}{\beta_{n+3}-}...\right],
 ~~(n=1,2...).
 \end{eqnarray}

The solution to the radial equation (\ref{ann}) is strictly
bounded up with the determination of the eigenvalue $A_{lm}$ of
the angular equation (\ref{12}), which appears explicitly in the
coefficients of the recurrence relation associated with the radial
equation. We could attempt to solve for both $\omega$ and $A_{lm}$
simultaneously, but this would be a time-consuming procedure.
Instead, we first fix the value of $a$, $l$, $m$ and $\omega$ and
find the angular separation constant $A_{lm}$ by looking for the
zero of the angular continued fraction (\ref{12}). Then, we use
the corresponding eigenvalue to look for the zeros of the radial
continued fraction. The n-th quasinormal frequency is
(numerically) the most stable root of the n-th inversion of the
continued fraction relation (\ref{ann}).

\section{Numerical results}

In this section we present the numerical results obtained by using
the numerical procedure just outlined in the previous section. The
results will be organized into three subsections: Dirac QNMs of
the Reissner-Nordstr\"om black hole, Dirac QNMs of the Kerr black
hole and Dirac QNMs of the Kerr-Newman black hole.

\subsection{Dirac QNMs of the Reissner-Nordstr\"om black hole}

As the angular momentum per unit mass  tends to zero
($a\rightarrow 0$) the Kerr-Newman metric reduces to the
Reissner-Nordstr\"om metric. In this limit the angular separation
constant $A_{lm}=\lambda^2=\left(l+\frac{1}{2}\right)^2$
\cite{Jing3} ( $l$ is the quantum number characterizing the
angular distribution). Thus, the QNMs in this static spacetime can
be obtained easily. The Dirac quasinormal frequencies of the
Reissner-Nordstr\"om black hole for $n=0,~1,~2,~3$ and $\lambda=1$
are listed in the table $I$ and that for $n=0,~1,~2,~3$ and
$\lambda=2$ are listed in the table $II$. From the tables we find
that both the real and imaginary parts of the quasinormal
frequencies decrease as the overtone number increases for the
fixed charge and angular quantum number (i.e., $Q=const.$ and
$l=const.$). We also find that both the real and imaginary parts
of the quasinormal frequencies increases as the angular quantum
number increases (except for the cases $Q=0.45$ and $n=0$) for the
fixed charge and overtone number.

\begin{table}
\caption{\label{tableRL1} Dirac quasinormal frequencies of the
Reissner-Nordstr\"om  black hole for $\lambda=1$.}
\begin{tabular}{c|c|c|c|c}
 \hline \hline
 ~~~$ Q $ ~~~ & $\omega$ ~($n=0$) & $\omega$~($n=1$) &
 $\omega$~($n=2$) & $\omega$~({$n=3$})\\
 \hline
0.00 &  0.365926-0.193965i & 0.295644-0.633857i &
0.240740-1.12845i  & 0.208512-1.63397i \\
0.05 &  0.366591-0.194066i & 0.296432-0.634048i &
0.241590-1.12866i  & 0.209400-1.63420i \\
0.10 &  0.368615-0.194366i & 0.298836-0.634601i &
0.244190-1.12923i  & 0.212121-1.63480i \\
0.15 &  0.372096-0.194851i & 0.302988-0.635442i &
0.248696-1.13001i  & 0.216843-1.63555i \\
0.20 &  0.377208-0.195488i & 0.309132-0.636419i &
0.255389-1.13068i  & 0.223871-1.63594i \\
0.25 &  0.384241-0.196213i & 0.317661-0.637242i &
0.264718-1.13064i  & 0.233661-1.63507i  \\
0.30 &  0.393651-0.196891i & 0.329196-0.637331i &
0.277339-1.12871i  & 0.246805-1.63118i \\
0.35 &  0.406184-0.197220i & 0.344692-0.635457i &
0.294100-1.12248i  & 0.263748-1.62065i  \\
0.40 &  0.423108-0.196443i & 0.365519-0.628639i &
0.315375-1.10611i  & 0.282749-1.59477i \\
0.45 &  0.446635-0.192202i & 0.391957-0.607831i &
0.333021-1.06224i  & 0.277400-1.52872i \\
\hline \hline
\end{tabular}
\end{table}

\begin{table}
\caption{\label{tableRL2} Dirac quasinormal frequencies of the
Reissner-Nordstr\"om  black hole for $\lambda=2$.}
\begin{tabular}{c|c|c|c|c}
 \hline \hline
 ~~~$ Q $ ~~~ & $\omega$ ~($n=0$) & $\omega$~($n=1$) &
 $\omega$~($n=2$) & $\omega$~({$n=3$})\\
 \hline
0.00 &  0.760074-0.192810i & 0.711666-0.595995i & 0.638523-1.03691i  & 0.569867-1.51490i \\
0.05 &  0.761372-0.192916i & 0.713058-0.595276i & 0.640042-1.03728i  & 0.571477-1.51529i \\
0.10 &  0.765327-0.193230i & 0.717299-0.596107i & 0.644677-1.03836i  & 0.576395-1.51643i \\
0.15 &  0.772119-0.193738i & 0.724597-0.597440i & 0.652672-1.04004i  & 0.584893-1.51813i \\
0.20 &  0.782084-0.194414i & 0.735335-0.599172i & 0.664476-1.04210i  & 0.597473-1.52001i \\
0.25 &  0.795772-0.195199i & 0.750138-0.601099i & 0.680818-1.04411i  & 0.614939-1.52137i  \\
0.30 &  0.814058-0.195973i & 0.770003-0.602799i & 0.702848-1.04523i  & 0.638525-1.52082i \\
0.35 &  0.838381-0.196464i & 0.796544-0.603349i & 0.732368-1.04363i  & 0.670048-1.51544i  \\
0.40 &  0.871259-0.195997i & 0.832478-0.600485i & 0.772116-1.03491i  & 0.711672-1.49828i \\
0.45 &  0.918762-0.192368i & 0.882263-0.587382i & 0.824382-1.00617i  & 0.760227-1.44942i \\
 \hline \hline
\end{tabular}
\end{table}

\subsection{Dirac QNMs of the Kerr black hole}

The Dirac quasinormal frequencies of the Kerr black hole for
$n=0,~1,~2,~3$ and $\lambda=1$ are listed in the table $III$ and
that for $n=0,~1,~2,~3$ and $\lambda=2$ are listed in the table
$IV$. We find that both the real and imaginary parts of the
quasinormal frequencies decrease as the overtone number increases
for the fixed angular momentum per unit mass  and angular quantum
number. We also find that both the real and imaginary parts of the
quasinormal frequencies increases as the angular quantum number
increases (except for the cases $a=0.4,~0.45$ and $n=0$) for the
fixed angular momentum per unit mass  and overtone number.

By comparing the results of the Reissner-Nordstr\"om and Kerr
black holes, it is interesting to note that the real part of the
quasinormal frequencies of the Reissner-Nordstr\"om black hole is
greater than that of the Kerr black hole, but the imaginary part
of the quasinormal frequencies of the Reissner-Nordstr\"om black
hole is less than that of the Kerr black hole if we take the same
values of the charge and the angular momentum per unit mass. That
is to say, the decay of the Dirac fields in the charged black-hole
spacetime is faster than that in the rotating black-hole spacetime
if the values of the charge and angular momentum per unit mass are
the same.

\begin{table}
\caption{\label{tableL1} Dirac quasinormal frequencies of the Kerr
black hole for $\lambda=1$.}
\begin{tabular}{c|c|c|c|c}
 \hline \hline
 ~~~$ a $ ~~~ & $\omega$ ~($n=0$) & $\omega$~($n=1$) &
 $\omega$~($n=2$) & $\omega$~({$n=3$})\\
\hline
0.00 &  0.365926-0.193965i & 0.295644-0.633857i & 0.240740-1.12845i  & 0.208512-1.63397i \\
0.05 &  0.366218-0.193805i & 0.296135-0.633141i & 0.241239-1.12700i  & 0.208875-1.63180i \\
0.10 &  0.367099-0.193316i & 0.297605-0.630951i & 0.242707-1.12257i  & 0.209884-1.62514i \\
0.15 &  0.368589-0.192461i & 0.300054-0.627144i & 0.245045-1.11486i  & 0.211268-1.61355i \\
0.20 &  0.370717-0.191175i & 0.303460-0.621458i & 0.248029-1.10334i  & 0.212433-1.59615i \\
0.25 &  0.373525-0.189350i & 0.307751-0.613449i & 0.251166-1.08710i  & 0.212101-1.57147i  \\
0.30 &  0.377060-0.186806i & 0.312708-0.602389i & 0.253306-1.06460i  & 0.207277-1.53691i \\
0.35 &  0.381352-0.183245i & 0.317679-0.587057i & 0.251404-1.03327i  & 0.189183-1.48777i  \\
0.40 &  0.386329-0.178141i & 0.320563-0.565425i & 0.234871-0.98938i  & 0.106512-1.41213i \\
0.45 &  0.391413-0.170595i & 0.313069-0.536419i & 0.178522-0.99702i  & 0.067565-1.44066i \\
 \hline \hline
\end{tabular}
\end{table}

\begin{table}
\caption{\label{tableL2} Dirac quasinormal frequencies of the Kerr
black hole for $\lambda=2$.}
\begin{tabular}{c|c|c|c|c}
 \hline \hline
 ~~~$ a $ ~~~ & $\omega$ ~($n=0$) & $\omega$~($n=1$) &
 $\omega$~($n=2$) & $\omega$~({$n=3$})\\
\hline
0.00 &  0.760074-0.192810i & 0.711666-0.594995i & 0.638523-1.03691i  & 0.569867-1.51490i \\
0.05 &  0.760593-0.192662i & 0.712351-0.594475i & 0.639418-1.03584i  & 0.570876-1.51315i \\
0.10 &  0.762165-0.192207i & 0.714416-0.592884i & 0.642106-1.03257i  & 0.573887-1.50779i \\
0.15 &  0.764829-0.191416i & 0.717897-0.590117i & 0.646595-1.02688i  & 0.578836-1.49851i \\
0.20 &  0.768656-0.190231i & 0.722849-0.585981i & 0.652877-1.01841i  & 0.585563-1.48468i \\
0.25 &  0.773753-0.188558i & 0.729342-0.580157i & 0.660882-1.00652i  & 0.593690-1.46530i  \\
0.30 &  0.780274-0.186246i & 0.737432-0.572132i & 0.670362-0.99020i  & 0.602309-1.43876i \\
0.35 &  0.788430-0.183045i & 0.747081-0.561063i & 0.680552-0.96784i  & 0.609060-1.40255i  \\
0.40 &  0.798494-0.178518i & 0.757886-0.545547i & 0.689090-0.93700i  & 0.607082-1.35381i \\
0.45 &  0.810750-0.171857i & 0.768079-0.523427i & 0.688731-0.89635i  & 0.580342-1.30276i \\
 \hline \hline
\end{tabular}
\end{table}

\subsection{Dirac QNMs of the Kerr-Newman black hole}

\begin{figure}
\includegraphics[scale=0.5]{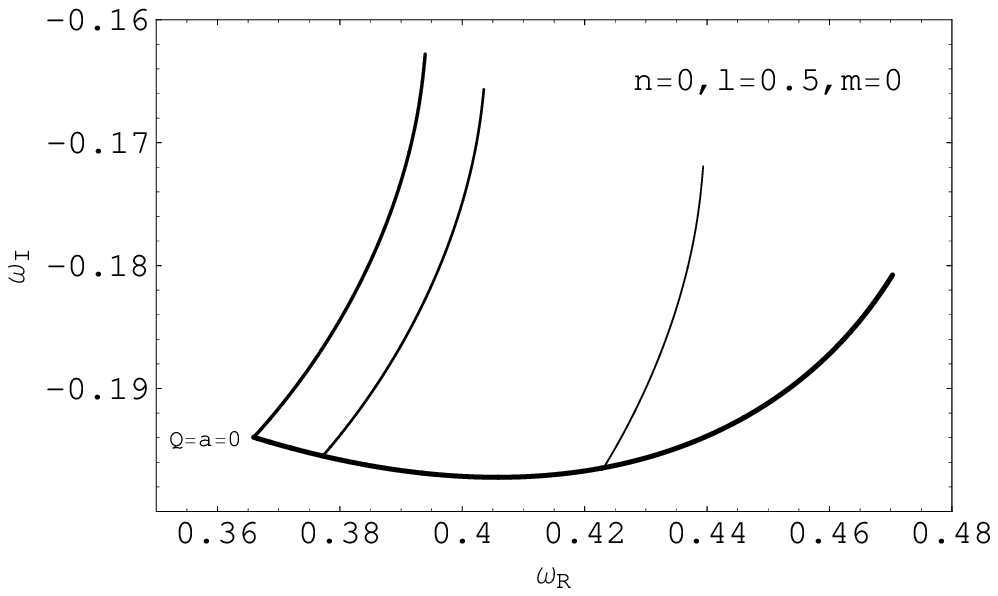}\hspace*{0.0cm}
\includegraphics[scale=0.5]{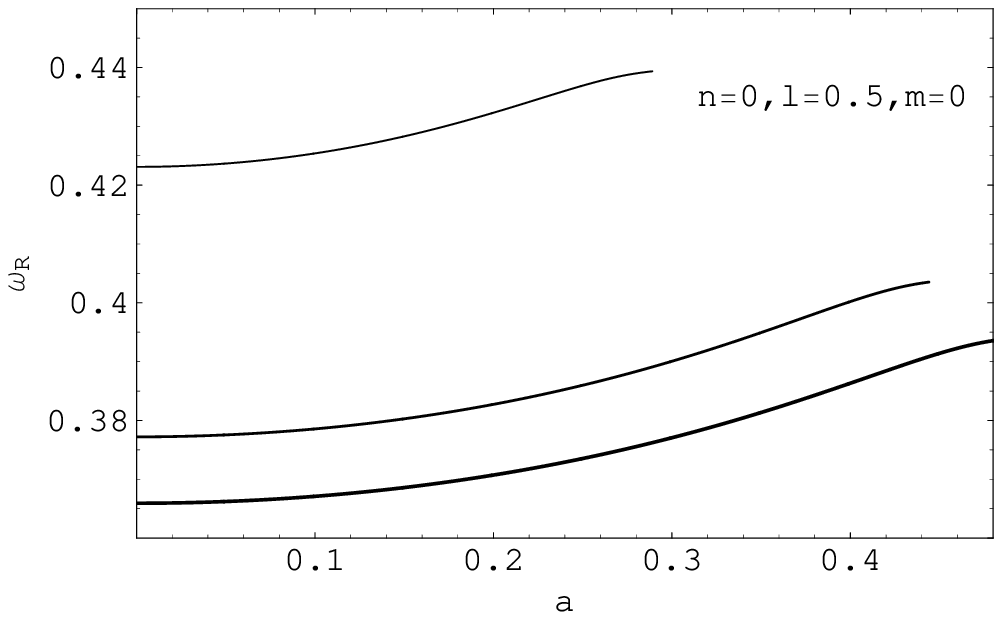}\hspace*{0.0cm}
\includegraphics[scale=0.5]{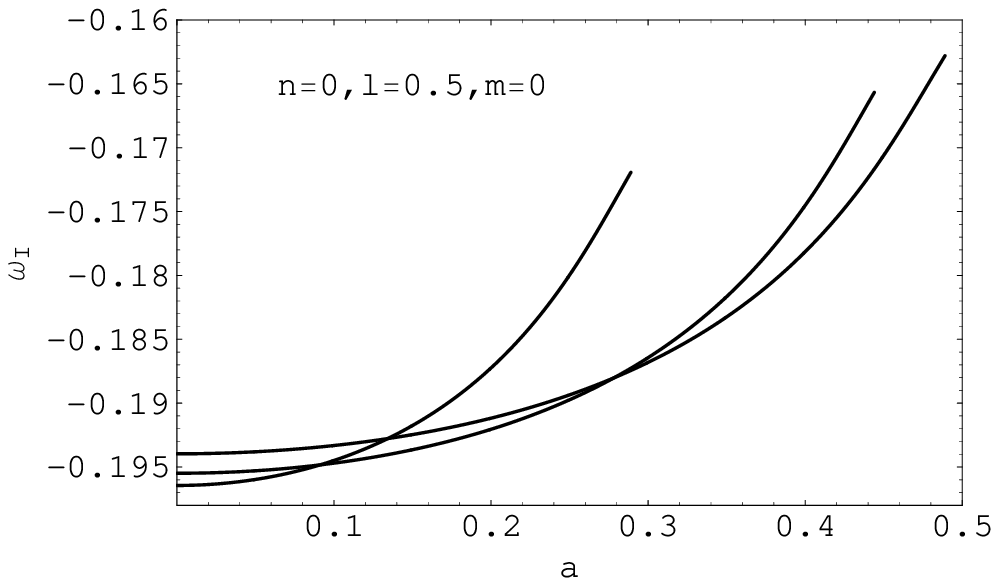}\\ \vspace*{0.5cm}
\includegraphics[scale=0.5]{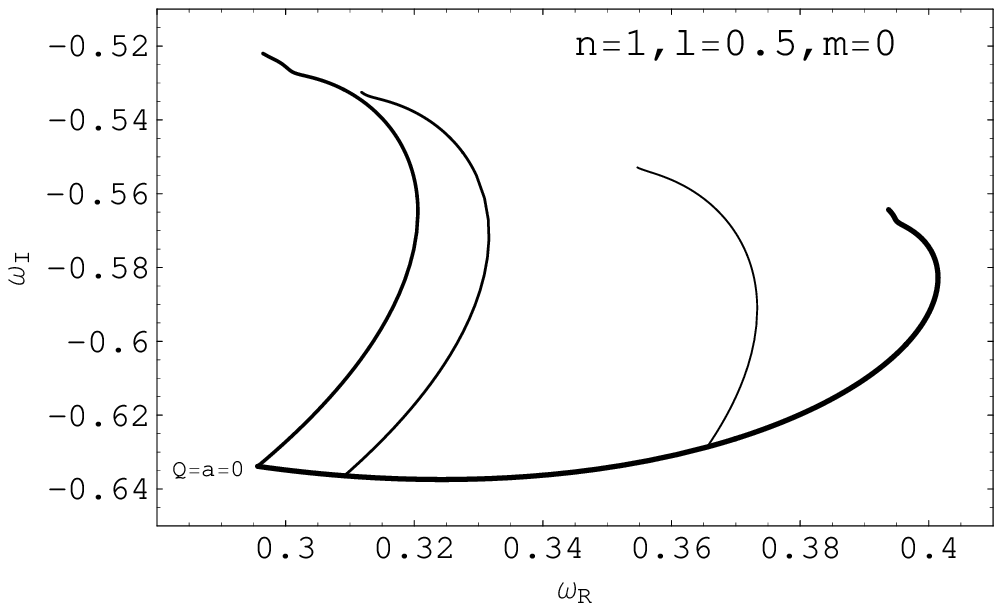} \hspace*{0.0cm}
\includegraphics[scale=0.5]{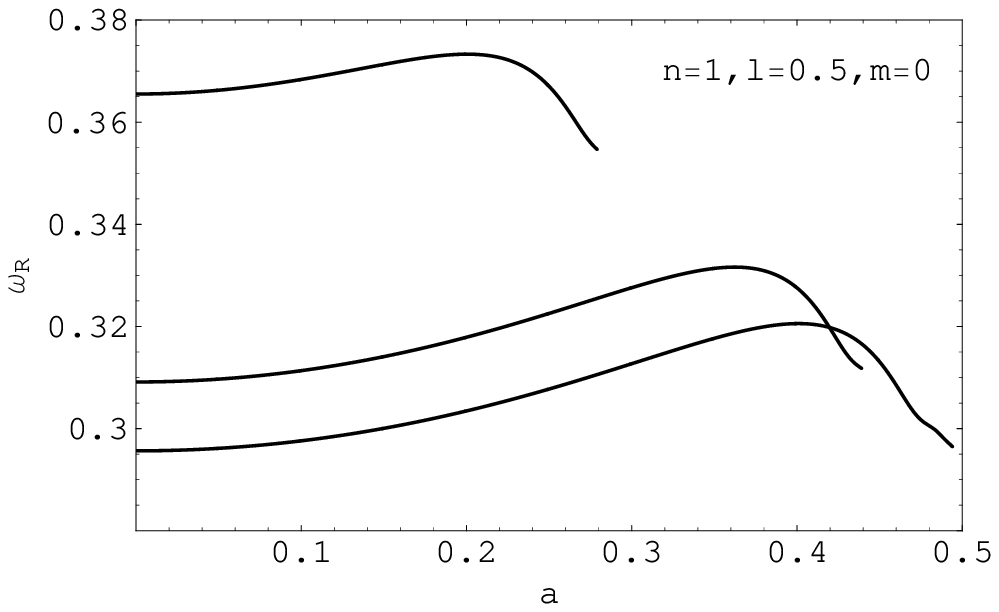} \hspace*{0.0cm}
\includegraphics[scale=0.5]{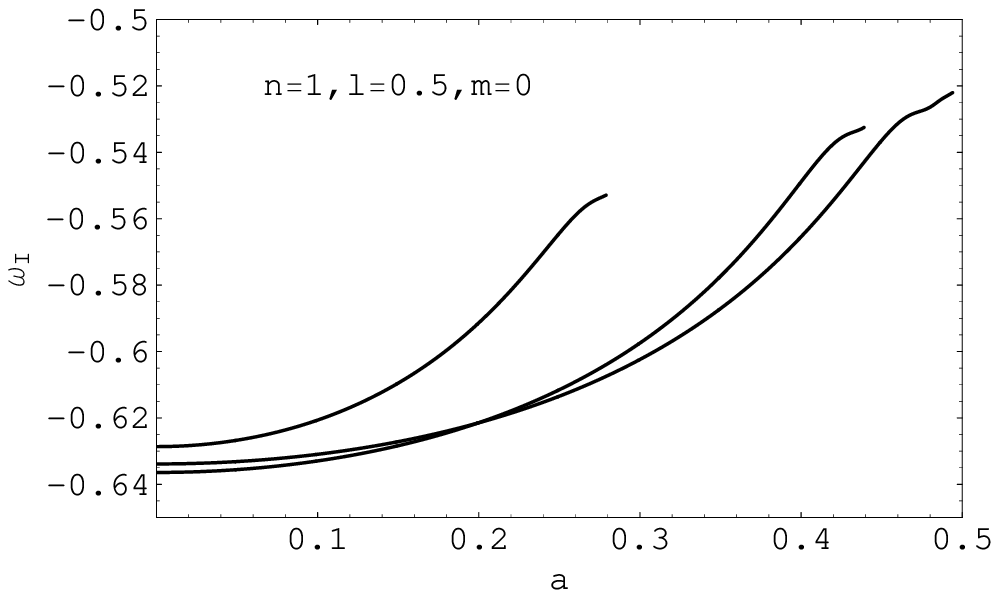}\\ \vspace*{0.5cm}
\includegraphics[scale=0.5]{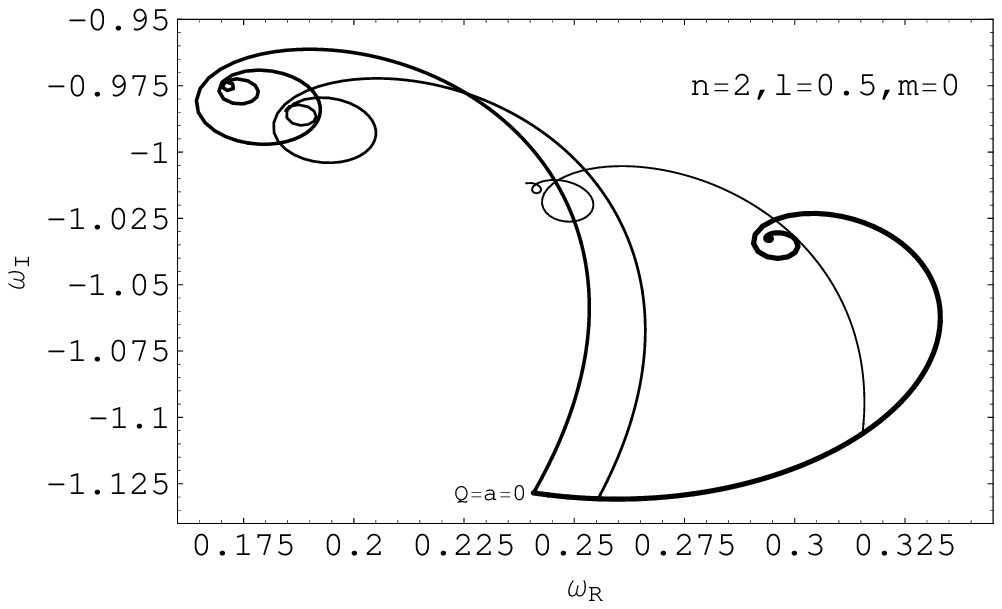} \hspace*{0.0cm}
\includegraphics[scale=0.5]{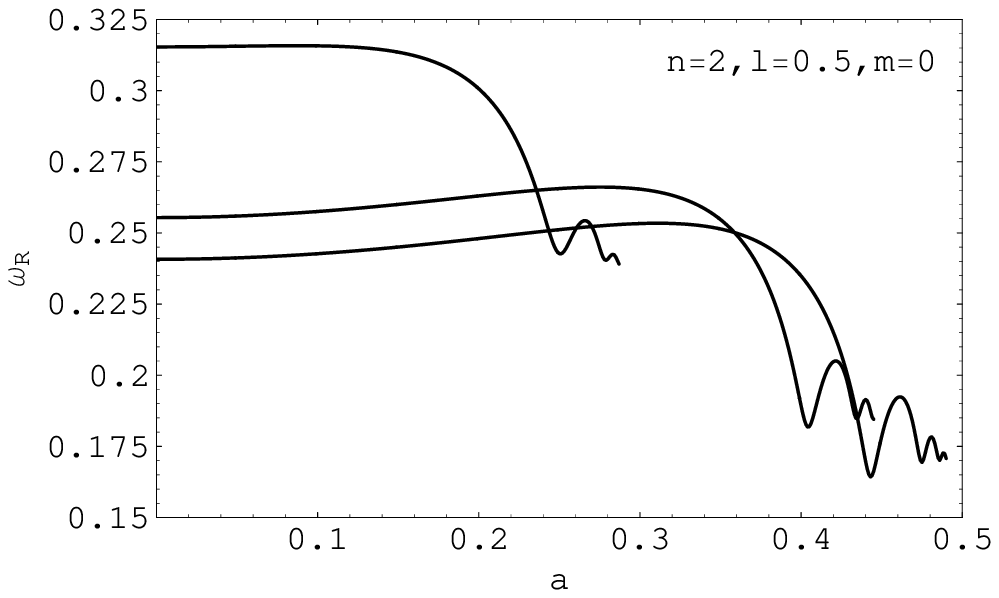} \hspace*{0.0cm}
\includegraphics[scale=0.5]{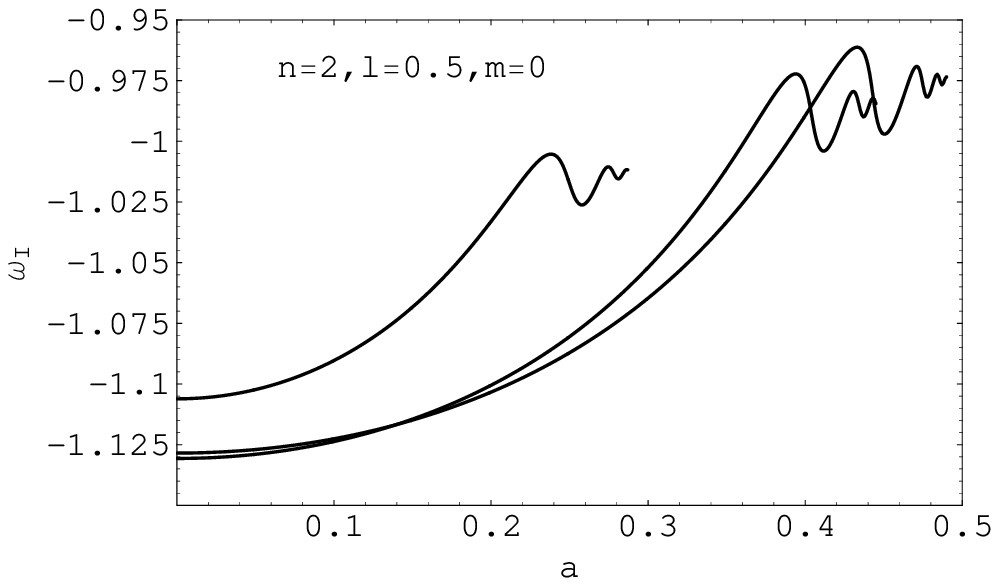}\\ \vspace*{0.50cm}
\caption{\label{fig1} The left three panels show trajectories in
the complex $\omega$ plane of the first three Dirac QNMs of the
Kerr-Newman black hole for $l=1/2$ and $m=0$. In each panel, the
thick curve corresponds to the Reissner-Nordstr\"om black hole
(a=0) which is drawn by increasing $Q$ from zero to the extremal
limit and the three thin curves are obtained by increasing $a$
from zero to the extremal limit for fixed values of the charge
$Q=0,~0.2,~0.4$ (left to right). These curves show that the
frequencies generally move counterclockwise as  $a$ (or $Q$)
increases and the number of spirals increases as the overtone
number increases. The other panels draw $Re(\omega)$ and
$Im(\omega)$ of the quasinormal frequencies versus $a$ in which
the curves from bottom to top refer to $Q=0,~0.2,~0.4$. These
panels tell us that, for $n\geq 2$, both the real and imaginary
parts are oscillatory functions of $a$ and the oscillations become
faster as the overtone number increases. }
\end{figure}

\begin{figure}
\includegraphics[scale=0.5]{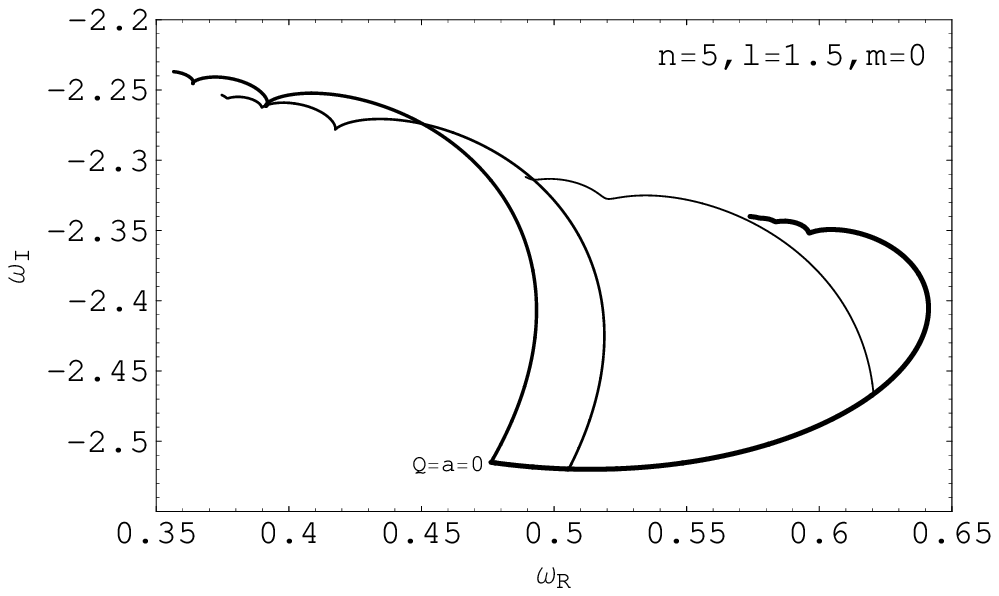}
\includegraphics[scale=0.5]{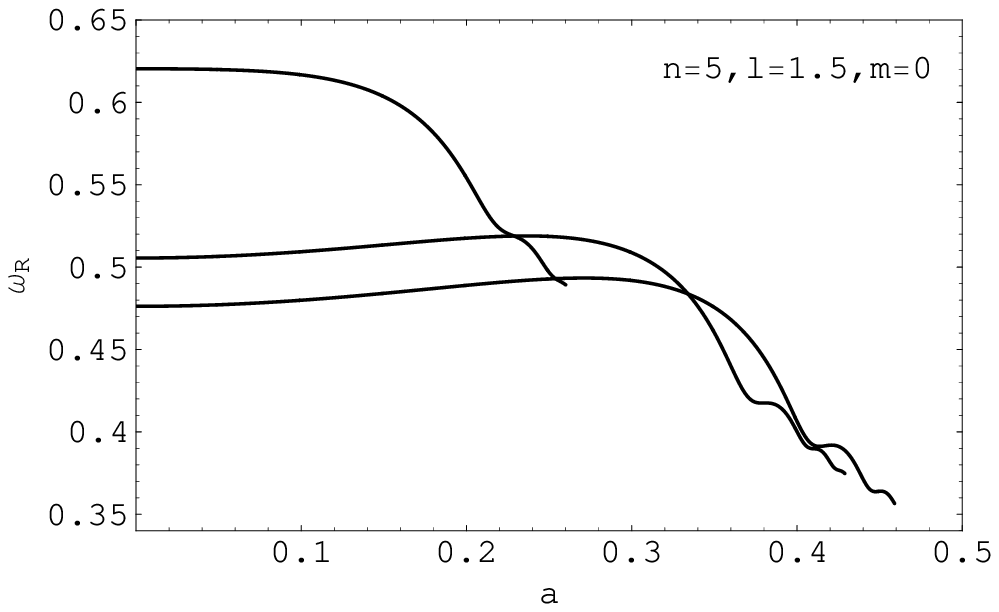}
\includegraphics[scale=0.5]{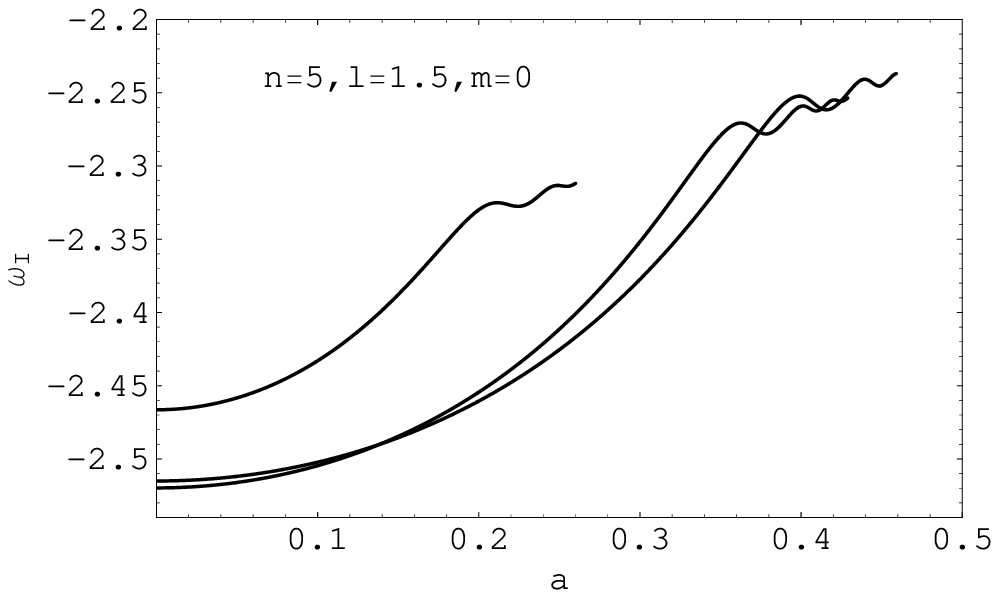} \\  \vspace*{0.5cm}
\includegraphics[scale=0.5]{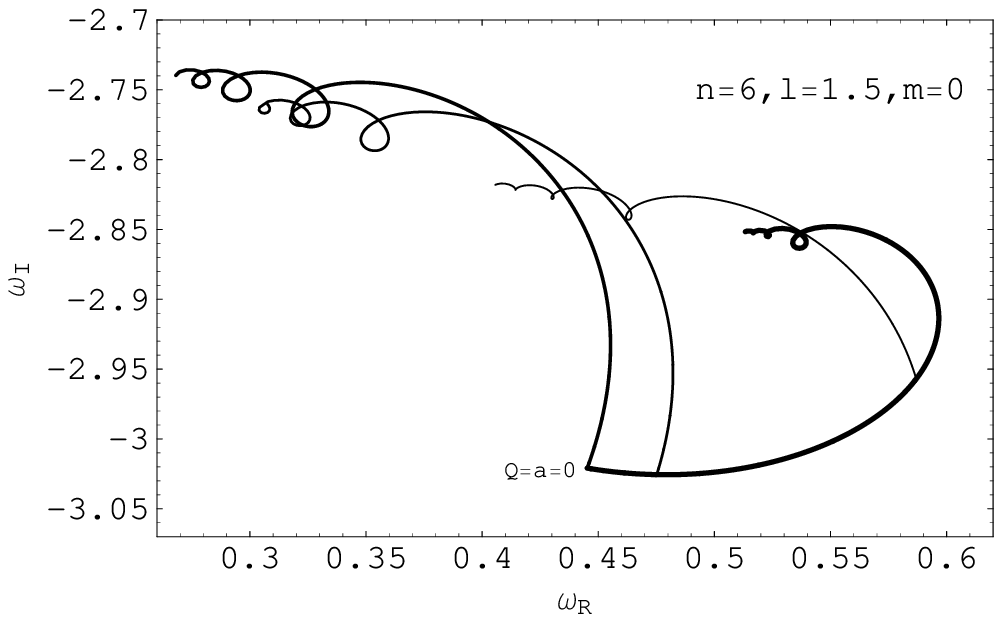}
\includegraphics[scale=0.5]{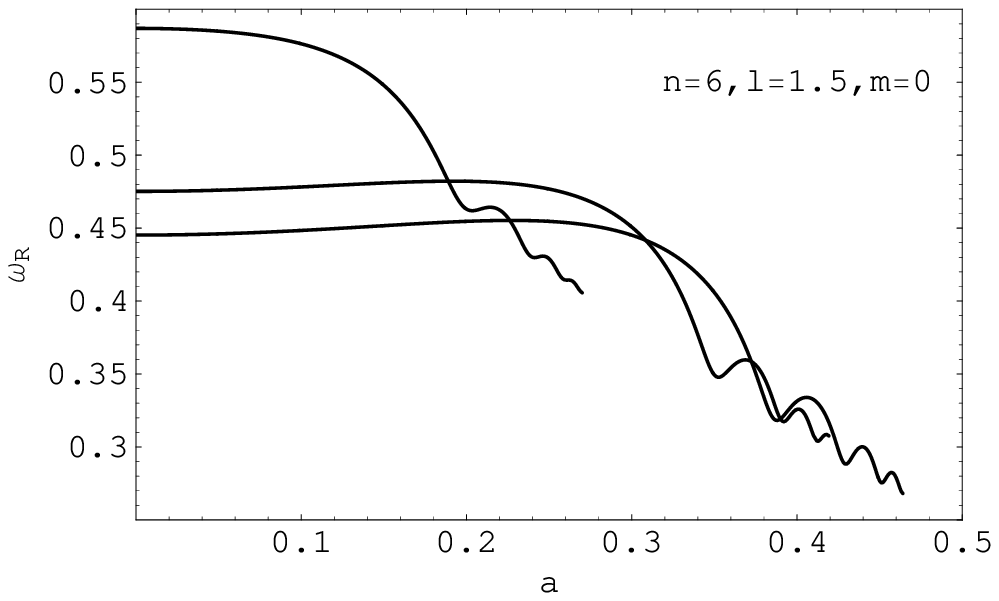}
\includegraphics[scale=0.5]{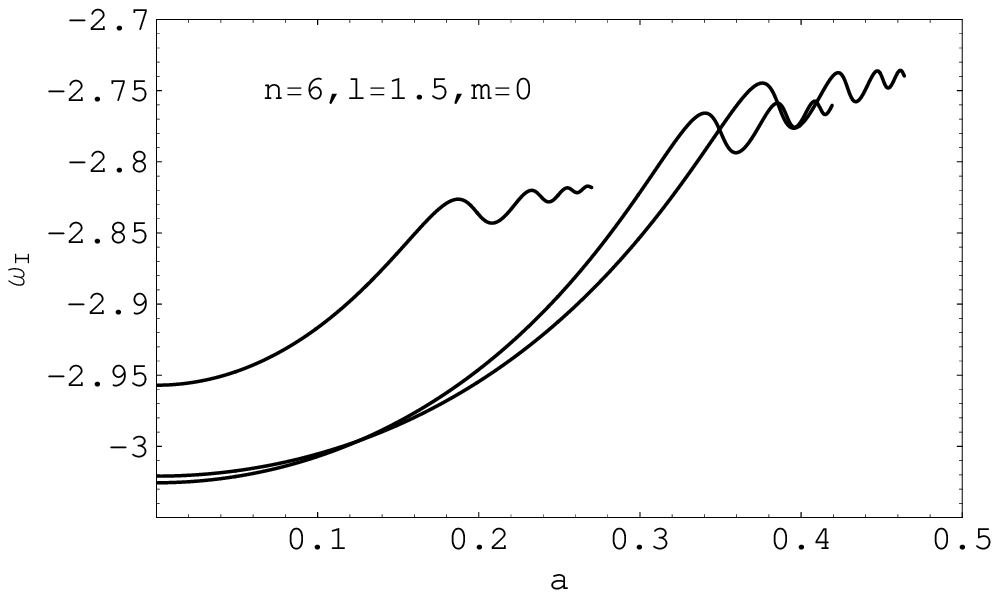} \\ \vspace*{0.5cm}
\includegraphics[scale=0.5]{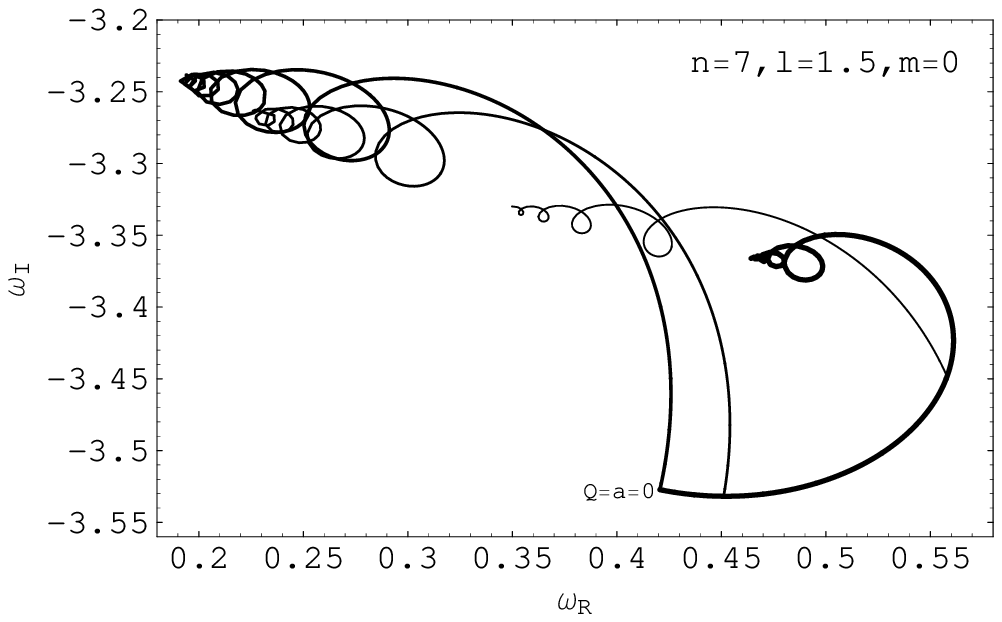}
\includegraphics[scale=0.5]{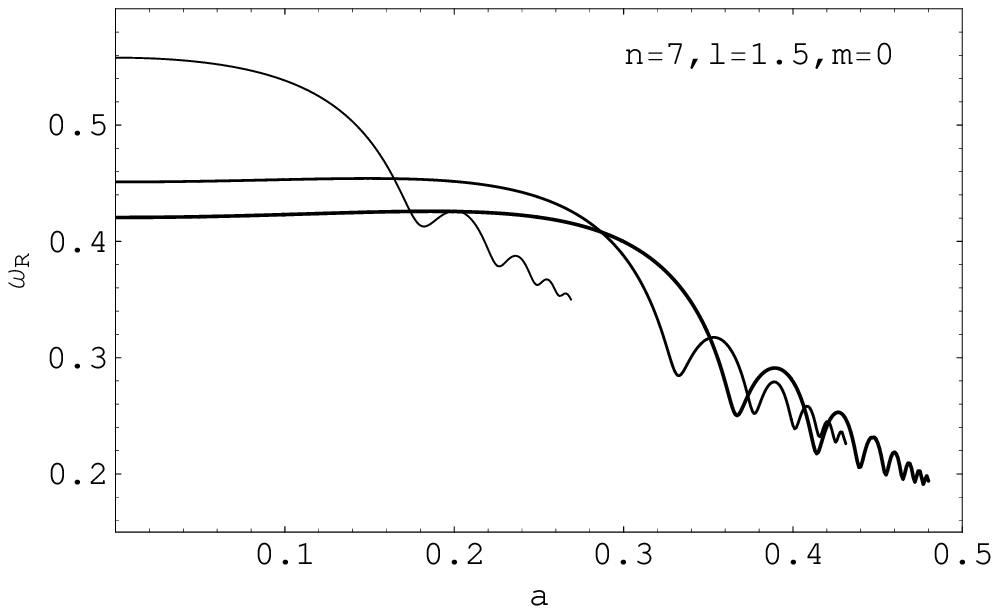}
\includegraphics[scale=0.5]{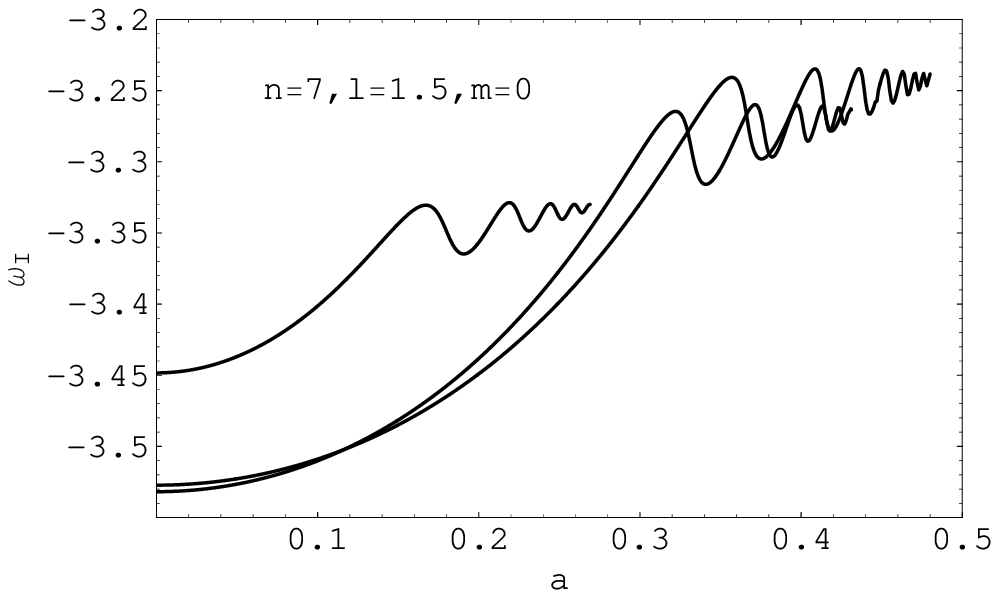}
\caption{\label{fig2} The left three panels describe the behavior
of the Dirac quasinormal frequencies of the Kerr-Newman black hole
for $n=5,~6,~7$, $l=3/2$ and $m=0$ in the complex $\omega$ plane.
In each panel, the thick curve is drawn by increasing $Q$ from
zero to the extremal limit for $a=0$ and the three thin curves are
obtained by increasing $a$ from zero to the extremal limit for
fixed values of the charge $Q=0,~0.2,~0.4$ (left to right). The
other panels draw $Re(\omega)$ and $Im(\omega)$ of the quasinormal
frequencies versus $a$ in which the curves from bottom to top
refer to $Q=0,~0.2,~0.4$. These panels tell us that, for $n\geq
5$, both the real and imaginary parts are oscillatory functions of
$a$ and the oscillations begin earlier as the charge increases.}
\end{figure}

The Dirac quasinormal frequencies of the Kerr-Newman black hole
for $n=0,~1,~2$ and $\lambda=1$  are shown by figure $1$ and those
for $n=5,~6,~7$ and $\lambda=2$  are shown by figure $2$. The left
columns in the Figs. \ref{fig1} and \ref{fig2} describe the
behavior of quasinormal frequencies in the complex $\omega$ plane
which show that the frequencies generally move counterclockwise as
the charge (the thick curves) or the angular momentum per unit
mass  (the thin curves) increases. They get a spiral-like shape,
moving out of the Schwarzschild  ($Q=0$ and $a=0$) or
Reissner-Nordstr\"om ($0<Q<0.5$ and $a=0$) values  and ``looping
in" towards some limiting frequencies as the charge and angular
momentum per unit mass  tend to the extremal value
$\sqrt{a^2+Q^2}=1/2$. For a given $\lambda$, we observe that the
number of spirals increases as the overtone number increases.
However, for a given overtone number $n$, the increasing $\lambda$
has the effect of ``unwinding" the spirals, as we see in the two
figures that the spiral begins at $n=2$ for $\lambda=1$ but
starts at $n=6$ for $\lambda=2$.

The second and last columns in the Figs. \ref{fig1} and \ref{fig2}
illustrate that the real and imaginary parts of the quasinormal
frequencies are the functions of the angular momentum per unit
mass  $a$. We know from these figures that both the real and
imaginary parts of the frequencies are oscillatory functions of
the angular momentum per unit mass. The oscillation starts earlier
and earlier as the overtone number $n$ grows for a fixed
$\lambda$, but it begins later and later as the angular quantum
number $a$ increases for a fixed $n$. We also learn that the
oscillation starts earlier and earlier as the charge $Q$ increases
for fixed $n$ and $\lambda$. Meanwhile, the oscillation becomes
faster as the overtone number increases for a given $\lambda$, but
it becomes slower as $\lambda$ increases for a given overtone
number $n$.

\section{summary}

The wave equations for the Dirac fields in the Kerr-Newman
black-hole spacetime are obtained by means of the Newman-Penrose
formalism. The expansion coefficients of the wave equation which
satisfy appropriate boundary conditions are directly determined by
a three-term recurrence relation. Then, the Dirac quasinormal
frequencies of the Kerr-Newman black hole are evaluated using
continued fraction approach and the results are presented by
tables and figures. By comparing the results of the
Reissner-Nordstr\"om and Kerr black holes, it is interesting to
note that the real part of the quasinormal frequencies of the
Reissner-Nordstr\"om black hole is greater than that of the Kerr
black hole, but the imaginary part is less than that of the Kerr
black hole if we take the same values of the charge and the
angular momentum per unit mass. That is to say, the decay of the
Dirac fields in the charged black-hole spacetime is faster than
that in the rotating black-hole spacetime if the values of the
charge and angular momentum per unit mass are the same. From
figures we find that the frequencies in the complex $\omega$ plane
generally move counterclockwise as the charge or the angular
momentum per unit mass  increases.  They get a spiral-like shape,
moving out of their Schwarzschild or Reissner-Nordstr\"om values
and ``looping in" towards some limiting frequencies as the charge
and angular momentum per unit mass tend to their extremal values.
For a given angular quantum number $\lambda$, we observe that the
number of spirals increases as the overtone number increases.
However, for a given overtone number, the increasing $\lambda$ has
the effect of ``unwinding" the spirals. We also find that the real
and imaginary parts of the quasinormal frequencies are oscillatory
functions of the angular momentum per unit mass, and the
oscillating behavior starts earlier and earlier as the overtone
number grows for a fixed $\lambda$ (or as the charge $Q$ increases
for fixed $n$ and $\lambda$), but it begins later and later as
$\lambda$ increases for a fixed overtone number.

\begin{acknowledgments}This work was supported by the
National Natural Science Foundation of China under Grant No.
10473004; the FANEDD under Grant No. 200317; and the SRFDP under
Grant No. 20040542003; the Hunan Provincial Natural Science
Foundation of China under Grant No. 04JJ3019; and the National
Basic Research program of China under Grant No. 2003CB716300.
\end{acknowledgments}

\end{document}